\documentclass[twocolumn,showpacs,amsmath,amssymb]{revtex4}

\usepackage[dvips]{graphicx}
\usepackage{color}

\def\setR{\mathbb{R}} 
 
\newcommand{\ket}[1]{\mid\!#1\,\rangle}

\newcommand{\sss}[1]{\scriptscriptstyle #1}

\begin{document}

\title{$SO(2,4)$-covariant quantization of the Maxwell field in conformally flat spaces}


\author{
Sofiane Faci}
\email{sofiane@cbpf.br}
\affiliation{
Institute of Cosmology, Relativity and Astrophysics (ICRA - CBPF)\\ 
Rua Dr. Xavier Sigaud, 150, CEP 22290-180, Rio de Janeiro, Brazil} 

\date{\today}

\begin{abstract}
We present an $SO(2,4)$-covariant quantization of the free electromagnetic field in conformally flat spaces (CFS). A CFS is realized  in a six-dimensional space as an intersection of the null cone with a given surface. The smooth move of the latter is equivalent to perform a Weyl rescaling. This allows to transport the $SO(2,4)$-invariant quantum structure of the Maxwell field from Minkowski space to any CFS. Calculations are simplified and the CFS Wightman two-point functions are given in terms of their Minkowskian counterparts. 
The difficulty due to gauge freedom is surpassed by introducing two auxiliary fields and using the Gupta-Bleuler quantization scheme.
The quantum structure is given by a vacuum state and creators/annihilators acting on some Hilbert space.
In practice, only the Hilbert space changes under Weyl rescalings. Also the quantum $SO(2,4)$-invariant free Maxwell field does not distinguish between two CFSs.

\end{abstract}

\maketitle

\section{Introduction}
The $SO(2,4)$-invariance of the Maxwell equation was discovered by Cunningham and Bateman a century ago. 
However in order to quantize the Maxwell field and due to gauge freedom, a gauge fixing condition is necessary.
The Lorenz gauge is usually used, which breaks the $SO(2,4)$ invariance. 
Nonetheless since such a symmetry mights apear to lack physical meaning, its breaking does not bother many people \cite{Kastrup}. The purpose of the present paper is to demonstrate the benefits of keeping this fundamental symmetry when quantizing the Maxwell field in conformally flat spaces (CFS). 

The starting point is the following. A classical $SO(2,4)$-invariant field cannot, at least locally, distinguish between two  CFSs \cite{Antoniadis:2011, Weinberg, Zumino, pconf7}. So why not maintain the $SO(2,4)$-invariance during the quantization process in a CFS?
Doing so, a free field living in a CFS might behave like in a flat space and the corresponding Wightman two-point functions can be related to their Minkowskian counterparts.
The work \cite{pconf3} confirms this assertion in the special case of Maxwell field in de Sitter space. Indeed, a new and simple  two-point Wightman function $\langle A_{\mu}(x) A_{\nu'}(x') \rangle$ was found and which has the same physical (gauge independent) content as the two-point function of Allen and Jacobson \cite{AllenJacobson}. This is because the Faraday propagator $\langle F_{\mu\nu}(x) F_{\mu'\nu'}(x') \rangle$ is the same.

The present work extends to general CFSs and clarify the quantum structure of the formalism developed in \cite{pconf3}. 
We use Dirac's six-cone formalism and realize all CFSs as intersections of the null cone and a given surface in a six-dimensional Lorentzian space. The introduction of auxiliary fields and the use of the Gubta-Bleuler quantization scheme are necessary to deal with gauge freedom of the Maxwell field. Another important ingredient is the use of a well-suited coordinate system. This allows to $SO(2,4)$-invariant CFS formulas to get a Minkowskian form.
 The main result is a set of Wightman two-point functions for Maxwell and auxiliary fields.

This paper is organized as follows. Sec. \ref{geom} sets the coordinates systems and the geometrical construction of CFSs.
Sec. \ref{fields} defines the fields and gives their dynamical equations.
In Sec. \ref{quantum-field}, the dynamical system is solved, the quantum field is explicitly constructed and the two-point functions are written down.  Some technical details are given in Appdx. \ref{details}.
The infinitesimal $SO(2,4)$ action on the fields $A_{\sss I}$ is expanded in  Appdx. \ref{action}  and their $SO(2,4)$-invariant scalar product is given in  Appdx. \ref{PS}.

\section{From $\setR^6$ to a four-dimensional CFS}\label{geom}
The six-dimensional Lorentzian space $\setR^6$ is provided with the natural orthogonal coordinates $y^\alpha = \{y^\mu, y^4,y^5\}$ and equipped with the metric $\tilde{\eta}_{\alpha \beta}=(+----+)$.  Quantities related to $\setR^6$ and its null cone $\mathcal{C}$ are labeled with a tilde.
We define a second coordinate system $x^{\sss I}=\{x^{c},x^{\mu},x^{+}\}$,
\begin{equation}\label{coord+muc}
 \begin{array}{lcllcl}
  x^c &=&   \dfrac{y_\alpha y^\alpha}{(y^4 + y^5)^2}   \ \qquad & 	y^\mu &=&x^+ \ x^\mu,\\
  x^\mu &=&  2 \dfrac{y^\mu}{y^4 + y^5} 			&	 y^4 &=&  x^+ (1  -x^c+  \frac{x^2}{4} ) \\
  x^+ &=&   \dfrac{ 1 }{2}(y^4 + y^5) 				&	 y^5 &=&  x^+(1  +x^c - \frac{x^2}{4} ) ,
 \end{array}
\end{equation}  
where the four components $x^{\mu}$ is the so-called polyspherical coordinate system \cite{Kastrup}
and $x^2= \eta_{\mu\nu}x^\mu x^\nu$.
A straightforward calculation yields
\begin{equation}\label{x+dx+}
y^{\alpha}\tilde{\partial}_{\alpha} = x^+ \partial_{+},
\end{equation}
which means that the component $x^+$ carries alone the homogeneity of the $y$'s. 
Using the system $x^{I}$, the null cone reads
\begin{equation}\label{C}
\mathcal{C} = \left\{x^c = 0 \right\}.
\end{equation}
A five-dimensional surface in $\setR^6$ is defined through
\begin{equation}\label{PK}
\mathcal{P}_{\sss \tilde K} = \left\{ x^+= \tilde K \right\},
\end{equation}
where the real and smooth function $\tilde K:=\tilde K(x^\mu, x^c)$ depends only on $x^{\mu}$ and $x^c$ and is then homogeneous of degree $0$. 
The intersection of $\cal C$ and  $\mathcal{P}_{\sss \tilde K}$ is a four-dimensional space
\begin{equation}\label{XK}
X_{\sss K} = \mathcal{C} \cap \mathcal{P}_{\sss \tilde K}.
\end{equation}
where the index $K$ in $X_{\sss K}$ refers to $K = \tilde K(x^\mu, x^c=0$).

\begin{figure}[h]
\includegraphics[width = 8cm]{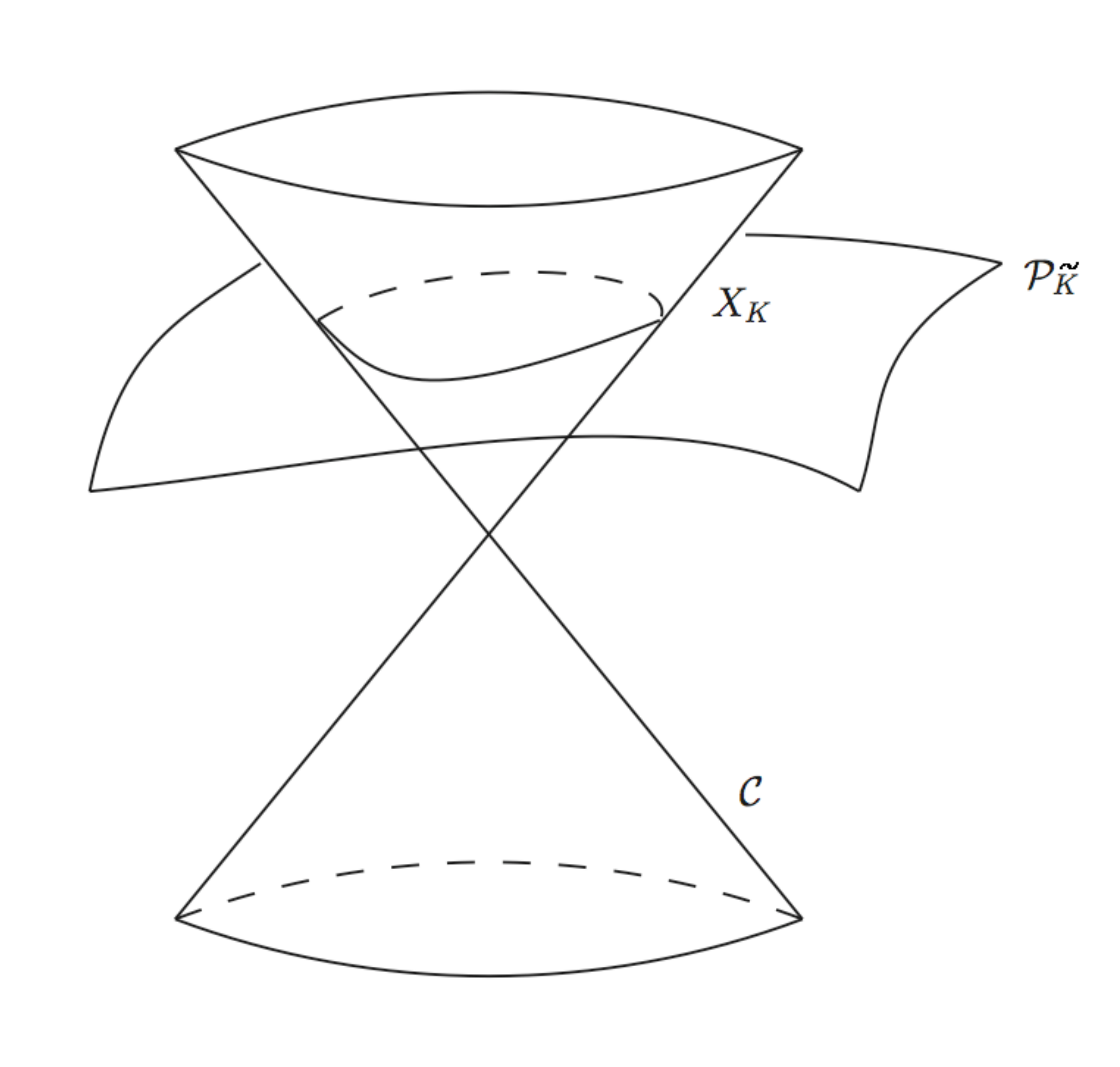}
\caption{The intersection of the null cone $\cal C$ with a five-dimensional surface ${\cal P}_{\sss \tilde K}$ produces a four-dimensional conformally flat space $X_{\sss K}$. 
}
\label{cone-ds}
\end{figure}

Regarding to its  metric inherited from $\tilde{\eta}_{\alpha\beta}$, precisely
\begin{equation}\label{lien-weyl}
ds_{\sss K}^2 = dy^{\alpha}dy_{\alpha}\bigg\vert_{\substack{x^c=0\\ x^+=\tilde K}} = K^2(x) \ \eta^{\mu\nu} dx_{\mu}dx_{\nu},
\end{equation}
 $X_{\sss K}$ turns out to be a CFS.
A smooth move of the surface ${\cal P}_{\sss \tilde{K}}$, which corresponds to changing the function $\tilde K$, amounts to perform a Weyl rescaling. This locally relates all CFSs and permits to go from one to another.
Note that  for $\tilde K=1$, $X_{\sss K}$ reduces to Minkowski space $X_{\sss M}$ and accordingly the four components system $x^{\mu}$ yields the usual cartesian system. 
The gradients
\begin{equation}\label{upsilon}
\tilde \Upsilon_{\sss I} = \frac{\partial}{\partial x^{\sss I}_{\sss K}} \ln \tilde K, 
\qquad \Upsilon_{\sss I} = \tilde \Upsilon_{\sss I} \bigg\vert_{\substack{x^c=0}},
\end{equation}
are extensivelly used in this article. 
The function $\tilde K$ does not depend on $x^+$ and thus $\Upsilon_+= \tilde \Upsilon_+ = 0$.

The choice of the function $\tilde K$,  including its $x^c$ dependence has to be done in such a way to ensure the invariance (in $\setR^6$) of the surface $\mathcal{P}_{\sss \tilde K}$ under the action of the isometry group associated to the desired $X_{\sss K}$ four-dimensional space.
Since the $\setR^6$ null-cone is $SO(2,4)$-invariant, the resulting $X_{\sss K}$ will be invariant under its isometry group.
Let us consider an example:
$\tilde K = [1+H^2(x^c - \frac{x^2}{4})]^{-1}$, 
where  $H^2$ is a constant. 
The associated surface $\mathcal{P}_{\sss \tilde K}$ and thus the corresponding $X_{\sss K}$ are left invariant under the action of de Sitter group \cite{pconf1}. Also, $X_{\sss K}$ is a de Sitter space.

\section{The Maxwell field from  $\setR^6$}\label{fields}
In this section, we explain how to obtain the $SO(2,4)$-invariant Maxwell field in $X_{\sss K}$ from a six-dimensional one-form. 
Following Dirac \cite{Dirac}, we consider a one-form $\tilde a$ defined in $\setR^6$ homogeneous of degree $r=0$ and which decomposes on the $\{dy^\alpha\}$ basis as
\begin{equation}\label{field-a}
\tilde a =\tilde a_\alpha dy^\alpha.
\end{equation}
The components $\tilde a_\alpha$ are
homogeneous of degree $r=-1$ and obey to the equation
\begin{equation}\label{equation-a}
\Box_{6} \ \tilde a_{\alpha} = \tilde \eta^{\beta\gamma} \frac{\partial}{\partial y^\beta}\frac{\partial}{\partial y^\gamma}  \ \tilde a_{\alpha}=0.
\end{equation}
This equation is naturally invariant under the $SO(2,4)$ action since this group has a linear action when acting in   $\setR^6$.
We then decompose the one-form $\tilde a$ on the basis $\{dx\}$ corresponding  to the system $x^{\sss I}$ (\ref{coord+muc}), with a slight but capital modification on the $dx^+$ component. There are two ways, the first decomposition reads 
\begin{equation}\label{eq-1}
\begin{split}
\tilde a & = \tilde{A}^{\sss M}_c dx^c + \tilde{A}^{\sss M}_\mu dx^\mu + \tilde{A}_+^{\sss M} \frac{dx^+}{x^+}.
\end{split}
\end{equation}
The second is given by
\begin{equation}\label{eq-2}
\begin{split}
\tilde a & = \tilde{A}^{\sss K}_c dx^c +\tilde{A}^{\sss K}_\mu dx^\mu + \tilde{A}_+^{\sss K} \frac{\tilde K}{x^+} d(\frac{x^+}{\tilde K}),
\\
& = (\tilde{A}^{\sss K}_c - \tilde \Upsilon_{c} \tilde{A}_+^{\sss K} ) dx^c 
+(\tilde{A}^{\sss K}_\mu - \tilde \Upsilon_{\mu} \tilde{A}_+^{\sss K} ) dx^\mu 
+ \tilde{A}_+^{\sss K} \frac{dx^+}{x^+}.
\end{split}
\end{equation}
Now, identifying (\ref{eq-1}) with (\ref{eq-2}), one obtains the relation between the fields $ \tilde{A}^{\sss K}_{\sss I}$ and $ \tilde{A}^{\sss M}_{\sss I}$ through  
\begin{equation}\label{extendedWeyl-bis-tilde}
\tilde A^{\sss K}_{\sss I} =\tilde  W_{\sss I}^{\sss J} \tilde A_{\sss I}^{\sss M} 
=\tilde    A_{\sss I}^{\sss M} + \tilde  \Upsilon_{\sss I} \tilde A_+^{\sss M}.
\end{equation}
All the fields $\tilde A_{\sss I}^{\sss K}$ and $\tilde A_{\sss I}^{\sss M}$ are by construction homogeneous of degree $r+1=0$. As a consequence, $\tilde A_{\sss I}^{\sss K}=A_{\sss I}^{\sss K}$ and $\tilde A_{\sss I}^{\sss M}=A_{\sss I}^{\sss M}$.
This amounts to project the fields $\tilde A_{\sss I}^{\sss K}$ on ${\cal P}_{\sss K}$ and $\tilde A_{\sss I}^{\sss M}$ on ${\cal P}_{\sss M}$. 
Then projecting the fields on the null cone $\cal C$ yields
\begin{equation}\label{extendedWeyl-bis}
 A^{\sss K}_{\sss I}  = W_{\sss I}^{\sss J} A_{\sss I}^{\sss M} 
 =   A_{\sss I}^{\sss M} + \Upsilon_{\sss I} A_+^{\sss M}.
\end{equation}
Thus $A^{\sss K}_{\sss I}$ and   $A^{\sss M}_{\sss I}$ are respectively  $X_{\sss K}$ and Minkowski fields.
Though in a slightly different maner,
this relation was obtained in \cite{pconf3} in the particular case of de Sitter space and was called the  ``extended Weyl transformation''. The fields $ A_{+}$ and $ A_{c}$ are auxiliary fields and the field $ A_{\mu}$ is, up to the condition  $ A_{+}=0,$  the Maxwell field. This will become clear here after.

Let us now turn to the dynamical equations. Our strategy is to transport  Minkowskian $SO(2,4)$-invariant equations to get  $SO(2,4)$-invariant equations in the $X_{\sss K}$ space. 
The first step is thus to write down the Minkowskian equations which are
 obtained using the equation (\ref{equation-a}) and the relation (\ref{a(A)-m}). This system reads
\begin{equation}\label{syst1M}
\left \{
\begin{aligned}
\partial^2 A_\mu^{\sss M} + \partial_\mu A_c^{\sss M} & =0 \\
\partial A^{\sss M} + A_c^{\sss M} & = \frac{1}{2} \partial^2  A_+^{\sss M}   \\
\partial^2 A_c^{\sss M} & = 0.
\end{aligned}
\right .
\end{equation}  
The corresponding system in $X_{\sss K}$ is obtained using (\ref{extendedWeyl-bis}),
\begin{equation}\label{syst1M-H}
\left \{
\begin{aligned}
\partial^2 A_\mu^{\sss K} + \partial_\mu A_c^{\sss K} &= \partial^2 ( \Upsilon_{\mu}A_{+}^{\sss K}) + \partial_{\mu} ( \Upsilon_{c}A_{+}^{\sss K})
\\
\partial A^{\sss K} +  A_c^{\sss K} & =  \left(  \Upsilon_{c} + \partial \Upsilon + \frac{1}{2} (\partial^2 + 2 \Upsilon \partial) \right) A_{+}^{\sss K}
 \\
\partial^2 A_c^{\sss H} & = \partial^2 ( \Upsilon_{c}A_{+}^{\sss K}),
\end{aligned}
\right .
\end{equation} 
where all  contractions are performed using $\eta_{\mu\nu}$ even though we are in the curved space $X_{\sss K}$.
The field $A_{\mu}$ obeying to the system above is not yet the Maxwell one.
Nevertheless, the constraint
$A_+^{\sss K}=0,$  
simplifies the system (\ref{syst1M-H})  and leads to 
\begin{equation}\label{MaxwellH1}
\left \{
\begin{aligned}
\partial^2 A_\mu^{\sss K} - \partial_\mu \partial  A^{\sss K} &=  0\\
\partial^2 \partial  A^{\sss K} &= 0.
\end{aligned}
\right .
\end{equation}
Despite their Minkowskian form, these equations are the Maxwell equation and a conformal gauge condition on any conformally flat space. This is due to the use of the polyspherical coordinate system (\ref{coord+muc}), which makes apparent the flatness feature of the $X_{\sss K}$ spaces.
The constraint $A+=0$ reduces the extended Weyl transformation (\ref{extendedWeyl-bis}) into  the identity 
\begin{equation}
{\displaystyle A^{\sss K}_I = A_I^{\sss M}},
\end{equation}
 recovering the ordinary vanishing conformal weight of the Maxwell field $A_{\mu}$. 
After some algebra, the covariant form of (\ref{MaxwellH1}) takes the form
\begin{equation}\label{system-covariant}
\left \{
\begin{aligned}
\Box A_\mu^{\sss K} - \nabla_\mu \nabla  A^{\sss K} + \frac{1}{4} R_{\mu\nu} A^{\nu}_{\sss K}  & = 0\\
 \nabla_{\mu} (\nabla^{\mu}\nabla^{\nu} + S^{\mu\nu}) A_{\nu}^{\sss K} & = 0,
\end{aligned}
\right .
\end{equation}
where $S_{\mu\nu} = -2 ( {R}_{\mu\nu} - \frac{1}{3} {R} g_{\mu\nu} )$.
The first line (resp. the second one) is the covariant Maxwell (resp. the Eastwood-Singer gauge \cite{EastwoodSinger}) equation in an arbitrary $X_{\sss K}$ space.
This conformal gauge was first derived by Bayen and Flato in Minkowski space \cite{bayen:1976}.
Its extension to curved spaces (even CFSs) is not trivial and can be performed using adapted tools like the Weyl-gauging technique \cite{Iorio:1996} or the Weyl-to-Riemann method \cite{pconf6}.

\newpage	
Note that the system (\ref{system-covariant}) is valid only if $A^+=0$ (an $SO(2,4)$-invariant constraint). But the latter has to be fixed at the end of the quantization process, not at the begining. Indeed, the auxiliary field $A_{+}$ acts as a Faddeev Popov ghost field and its retention during the quantization process is necessary. The constraint $A^+=0$ will be applied on the quantum space to select an invariant subspace of physical states and the Wightman functions thus include the whole big space.
This is related to the undecomposable group representation 
 (see appendix \ref{action}).

\section{Canonical quantization}\label{quantum-field}
We now apply the Gupta-Bleuler quantization scheme \cite{Gupta, Bleuler, grt}. This can be summarized as follows. 
We have seen that $A^{\sss K}_\mu$ is interpreted as the Maxwell field in the Eastwood-Singer gauge (\ref{system-covariant}) on the space $X_{K}$ when the constraint $A_+=0$ is applied. 
The problem is that pure gauge solutions ($A^{\sss K}_\mu = \nabla_\mu\Lambda$, with 
$ \nabla_{\mu} (\nabla^{\mu}\nabla^{\nu} + S^{\mu\nu})  \nabla_{\nu} \Lambda = 0$ and $A_+=0$)  are orthogonal to all the solutions including themselves. As a consequence, the space of solutions is degenerate and no Wightman functions can be constructed.
To fix this problem, we consider the system (\ref{syst1M-H}), instead of (\ref{system-covariant}), 
for which $A_+\neq 0$ and thus a causal reproducing kernel can be found.
This means that for quantum fields $\hat A_{\sss I}$ acting on some Hilbert (or Krein) space $\cal H$, we cannot impose the operator equation $\hat A_{\sss +}=0$. Instead, we define the subspace of  physical states $\mathcal{H}_{phy}\in\cal H$ which cancels the action of $\hat A^+$. Then the Maxwell equation and the Eastwood-Singer gauge hold in the mean on the space $\mathcal{H}_{phy}$. 
The task seems complicated at first sight, but thanks to the correspondence (\ref{extendedWeyl-bis}) we only need to solve the Minkowskian system (\ref{syst1M}), which is already done in \cite{Bayen}. Indeed, using the Weyl equivalence between CFSs, the whole structure of an $SO_{\sss O}(2,4)$-covariant free field theory can be transported from Minkowski to another CFS.
In the following, we solve the dynamical equations, obtain the modes, determine the quantum field, the subspace of physical states and finally compute the two-point functions.

\subsection{The mode solutions}\label{solutions}
The solutions of the Minkowskian system (\ref{syst1M}) can be obtained from \cite{pconf3} and read
\begin{equation}\label{modes-A}
A^{\sss M}_{\sss lm (\alpha)I}(x) = \epsilon^{\sss M}_{\sss (\alpha)I}(x) \ \phi^{\sss M}_{\sss lm}(x),
\end{equation}
where $\epsilon^{\sss M}_{\sss (\alpha)}(x)$ are polarization vectors whose components are given by $\epsilon^{\sss M}_{\sss (\alpha)I} (x)= S_{\sss I}^{\sss \beta} (x)\ \tilde\eta_{\sss \alpha\beta}$ and verifying 
$\langle \epsilon^{\sss M}_{\sss (\alpha)}, \epsilon^{\sss M}_{\sss (\beta)} \rangle = - \tilde \eta_{\alpha\beta},
$ 
with respect to the scalar product (\ref{scalarPS-A}).
 The matrix $S_{\sss I}^{\sss \beta}(x)$ relates the fields $A_{\sss I}^{\sss M}$ and $a_{\sss \alpha}^{\sss M}$ (\ref{matrix-S}). The scalar modes $\phi^{\sss M}_{\sss lm}(x)$ are solutions of the Minkowskian $SO(2,4)$-invariant (or massless) sclalar field equation $\partial^2 \phi^{\sss M} = 0$,
\begin{equation}\label{modes-phi}
\phi^{\sss M}_{\sss lm}(x) =  c_{l} \ \mathcal{Y}_{lm}(x),
\end{equation}
where $\mathcal{Y}_{lm}(x)$ denotes the usual hyperspherical harmonics. The normalization constant $c_{l}$ is chosen in order to get
$\langle \phi^{\sss M}_{\sss lm}, \phi^{\sss M}_{\sss l'm' }\rangle_{\sss KG} = \delta_{\sss ll'}\delta_{\sss mm'},$
with respect to the Klein-Gordon scalar product. 
As a consequence, the solutions (\ref{modes-A}) are normalized with respect to (\ref{scalarPS-A}),
\begin{equation}\label{norm}
\langle A^{\sss M}_{{\sss lm (\alpha)}}, A^{\sss M}_{{\sss l'm' (\beta)}}\rangle = - \tilde\eta{\sss \alpha\beta}\delta_{\sss ll'}\delta_{\sss mm'}.
\end{equation} 
Thus the general solution of the system (\ref{syst1M}) reads
\begin{equation}\label{generalsolution-AM}
A^{\sss M}_{}(x)=\sum_{{\sss lm\alpha}} a_{\sss lm (\alpha)}^{\sss M} A^{\sss M}_{{\sss lm (\alpha)}}(x),
\end{equation}
where $a_{\sss lm (\alpha)}^{\sss M}$ are real constants.

Let us now turn to the modes of the system (\ref{syst1M-H}). They are obtained thanks to the extended Weyl transformation (\ref{extendedWeyl-bis}) applied on the Minkowskian modes (\ref{modes-A})
\begin{equation}\label{modes-A-K}
A^{\sss K}_{\sss lm (\alpha)I}(x) = A^{\sss M}_{\sss lm (\alpha)I}(x) + \Upsilon_{\sss I}(x) A^{\sss M}_{\sss lm (\alpha)+}(x).
\end{equation}
These modes are normalized like (\ref{norm}) but according to the scalar product (\ref{scalarPS-A-K}).
The general solution on $X_{K}$ reads
\begin{equation}\label{generalsolution-AK}
A^{\sss K}_{}(x)=\sum_{{\sss lm\alpha}} a_{\sss lm (\alpha)}^{\sss K} A^{\sss K}_{{\sss lm (\alpha)}}(x).
\end{equation}
where the $a_{\sss lm (\alpha)}^{\sss K}$ are some real constants.

Note that when 
\begin{equation}\label{A+=0}
A_+^{\sss M}=A_+^{\sss K}=A_{+}= 0,
\end{equation}
the solutions (\ref{generalsolution-AM}) and (\ref{generalsolution-AK}) solve the Maxwell equation in the Eastwood-Singer gauge.

\subsection{The quantum field and physical states}
We can now define the quantum fields and construct  the Fock spaces as usual. 
The quantum fields corresponding to (\ref{generalsolution-AM}) and (\ref{generalsolution-AK}) are respectively defined through 
\begin{equation}
\hat{A}^{\sss M}_{}(x)=\sum_{{lm\alpha}} A^{\sss M}_{\sss lm (\alpha)}(x)
 \hat a_{\sss lm (\alpha)} + A^{{\sss M}\,*}_{\sss lm (\alpha)}(x) \hat a^{\dag}_{\sss lm (\alpha)}, \label{q-field-M}
\end{equation}
\begin{equation}
\hat{A}^{\sss K}_{}(x)=\sum_{lm\alpha} A^{\sss K}_{\sss lm (\alpha)}(x)
\hat a_{\sss lm (\alpha)} + A^{{\sss K}\,*}_{\sss lm (\alpha)}(x) \hat a^{\dag}_{\sss lm (\alpha)},\label{q-field}
\end{equation}
where the operators $\hat a_{\sss lm (\alpha)}$ and $\hat a_{\sss lm (\alpha)}^\dag$
are respectively the annihilators and creators of the modes (\ref{modes-A}) in $X_{\sss M}$ and the modes 
(\ref{modes-A-K}) in $X_{\sss K}$.
The use of the same annihilators and creators for all CFSs is highly  important for our purpose. Indeed, this allows to define the the same vaccuum state $ \ket{0}$ through
\begin{equation}\label{vacuum}
\hat a_{\sss lm (\alpha)}\  \ket{0} = 0,
\end{equation}
for any annihilator. 
The one-particle states are built by applying the  creators on the vacuum state
\begin{equation}\label{quantum-state}
\ket{ A_{\sss lm (\alpha)}} = \hat a_{\sss lm (\alpha)}^{\dag} \ket{0}.
\end{equation}
and the multiple particle states of the Fock spaces are constructed as usual.
Moreover, the  annihilation and creation operators obey to the following algebra
\begin{equation}
\begin{split}
& [  \hat a_{\sss lm (\alpha)} , \hat a_{\sss l'm' (\beta)}] =   [  \hat a_{\sss lm (\alpha)}^{\dag} , \hat a_{\sss l'm' (\beta)}^{\dag}] = 0
\\
& [  \hat a_{\sss lm (\alpha)} , \hat a_{\sss l'm' (\beta)}^{\dag}]  = -\tilde \eta_{\sss \alpha\beta} \delta_{\sss ll'} \delta_{\sss mm'}.
\end{split}
\end{equation}

The subset of physical states in both spaces is determined thanks to the classical physical solutions  (\ref{generalsolution-AM}) and (\ref{generalsolution-AK}) verifying (\ref{A+=0}).
In quantum language,
$\ket{A_{phy}}$ is a physical state iff
\begin{equation}\label{physicalstateDef}
\hat{A}^{\sss (+)}_+ \ket{A_{phy}} = 0,
\end{equation}
where $\hat{A}^{\sss (+)}_+ $ is the annihilator part of $\hat{A}_+$. 
This implies the equality 
\begin{equation}
\langle A_{phy} \mid \hat{A}_+ (x)\mid B_{phy} \rangle =0,
\end{equation}
for any physical states $\ket{A_{phy}}$ and $\ket{B_{phy}}$. 
Also, the subspace of physical states is the same in all CFSs, which allows to transport physical quantities from Minkowski space into the $X_{\sss K}$ space.
As a consequence, one obtains
\begin{equation}
\begin{cases}
\langle A_{\sss phy} \mid  \Box \hat A_\mu^{\sss K} - \nabla_\mu \nabla  \hat A^{\sss K} + \frac{1}{4} R_{\mu\nu} \hat A^{\nu}_{\sss K}    \mid B_{\sss phy} \rangle =0\\
\langle A_{\sss phy} \mid \qquad  \nabla_{\mu} (\nabla^{\mu}\nabla^{\nu} + S^{\mu\nu}) \hat A_{\nu}^{\sss K} 
\quad \mid B_{\sss phy}\rangle =0,
\end{cases}
\end{equation}
in $X_{\sss K}$ and the corresponding minkowskian system in $X_{\sss M}$.
The quantum fields fulfill the Maxwell equation together with the Eastwood-Singer gauge in the mean on the physical states.

\subsection{Two-point functions}\label{2points}
We show in this part how to get the Wightman two-point functions on $X_{\sss K}$ from their Minkowskian counterparts. 
The Wightman functions related to the Minkowskian fields $A_{\sss I}^{M}$ are defined through
\begin{equation}\label{MaxweldS2ptDef}
\begin{split}
D^{\sss M}_{\sss IJ'}(x,x') & = \langle0 \mid \hat{A}^{\sss M}_{\sss I}(x)\hat{A}^{\sss M}_{\sss J'}(x') \mid 0 \rangle, 
\\ & =\sum_{lm\alpha} A^{\sss M}_{\sss lm (\alpha) I}(x)\ A^{\sss M}_{\sss lm (\alpha) J'}(x').
\end{split}
\end{equation}
Their expressions are given in \cite{Bayen} and read
\begin{equation}\label{D-mink}
\begin{cases}
D^{\sss M}_{\mu\nu'}(x,x') & = + \eta_{\mu\nu'}  \ D^s_{\sss M}(x,x')
\\
D^{\sss M}_{\mu+}(x,x') & = -  (x-x')_{\mu}  \ D^s_{\sss M}(x,x')
\\
D^{\sss M}_{c+}(x,x') & = -  2 \ D^s_{\sss M}(x,x')
\\
D^{\sss M}_{++}(x,x') & =  + \frac{1}{8 \pi^2}
\\
D^{\sss M}_{\mu c}(x,x') & =  0
\\
D^{\sss M}_{c c}(x,x') & =  0,
\end{cases}
\end{equation}
where 
$D^s_{\sss M}(x,x')=\frac{-1}{8 \pi^2 \sigma_{0}},$ 
 with $\sigma_{0}=\frac{(x-x')^2}{2},
 $
stands for the Wightman two-point function related to the Minkowskian massless scalar field.

The Wightman two-point functions related to the field $ A^{\sss K}_{\sss I}$ are given by \begin{equation}\label{wightman-K}
\begin{split}
D^{\sss K}_{\sss IJ'}(x,x') &= \langle0 \mid \hat{A}^{\sss K}_{\sss I}(x)\hat{A}^{\sss K}_{\sss J'}(x') \mid 0 \rangle,
\\ & =\sum_{lm\alpha} A^{\sss K}_{\sss lm (\alpha) I}(x)\ A^{\sss K}_{\sss lm (\alpha) J'}(x').
\end{split}
\end{equation}
Now, using (\ref{modes-A-K}), (\ref{MaxweldS2ptDef}) and (\ref{wightman-K}), allows to write the following capital formula 
\begin{equation}\label{DK-DM}
{\displaystyle  D^{\sss K}_{IJ'}(x,x') = D^{\sss M}_{IJ'}(x,x') + \Theta^{\sss K}_{IJ'}(x,x')  },
\end{equation}
where the $\Theta$ terms read 
\begin{equation}
\begin{split}
 \Theta_{\sss IJ'}^{\sss K}(x,x') 
= & \Upsilon_{\sss I}(x) D^{\sss M}_{+J'}(x,x') 
\\& 
+ \Upsilon_{\sss J'}(x') D^{\sss M}_{I+}(x,x') 
\\& 
 + \Upsilon_{\sss I}(x)\  \Upsilon_{\sss J'}(x') \ D^{\sss M}_{++}(x,x').
\end{split}
\end{equation}
%
The Wightman two-point functions (\ref{DK-DM}) read
\begin{equation}\label{D-X-K-relevent}
\begin{cases}
D^{\sss K}_{\mu\nu'} & =  + \Big( \eta_{\mu\nu} -  \Upsilon_{\mu} \Upsilon'_{\nu'} \ \sigma_{0} 
 - \Upsilon_{\mu} (x-x')_{\nu'} 
 \\& \qquad \qquad \qquad
  - \Upsilon'_{\nu'} (x-x')_{\mu} \Big) D^s_{\sss M}
\\
D^{\sss K}_{\mu+} & = -\Big(  \Upsilon_{\mu} \sigma_{\sss 0} + (x-x')_{\mu}   \Big) D^s_{\sss M}
\\
D^{\sss K}_{c+} & = - \Big(  2 + \Upsilon_{c} \sigma_{0}  \Big) D^s_{\sss M}
\\
D^{\sss K}_{++} & =  + \, \frac{1}{8 \pi^2}
\\
D^{\sss K}_{\mu c}  & = - \Big(  \Upsilon_{\mu} \Upsilon'_{c} \ \sigma_{\sss 0}  
 + 2 \Upsilon_{\mu} + \Upsilon'_{c} (x-x')_{\mu}   \Big)  D^s_{\sss M}
\\
D^{\sss K}_{c c} & = - \Big( \Upsilon_{c} \Upsilon'_{c} \ \sigma_{\sss 0}  
+ 2 \Upsilon_{c} + 2 \Upsilon'_{c}   \Big) D^s_{\sss M},
\end{cases}
\end{equation}
where $D_{\sss IJ'}^{\sss K}\equiv D^{\sss K}_{\sss IJ'}(x,x')$, 
$D_{s}^{\sss M}\equiv D_{s}^{\sss M}(x,x')$, $\Upsilon_{\sss I}\equiv\Upsilon_{\sss I}(x)$ and $\Upsilon'_{\sss I} \equiv\Upsilon_{\sss I}(x')$.
\\
 
 To end this paper, let us consider an important particular case, that corresponding to de Sitter space. This case is obtained by specifying 
$$\tilde K= \frac{1}{1+H^2(x^c - \frac{x^2}{4})}, 
$$
where $H$ is related to the de Sitterian Ricci scalar through $R=12H^2$. 
The gradients (\ref{upsilon}) read
$$
\Upsilon_{\mu} = \frac{H^2K}{2} \eta_{\mu\nu} x^{\nu}, \quad \Upsilon_{c} = -\frac{H^2K}{2}.
$$
In this case we obtain simple expressions for the two-point functions related to the fields $A_{\sss I}^{H}$ on de Sitter space. The three more relevant yield
\begin{equation}\label{D-ds}
\begin{cases}
D^{\sss H}_{\mu\nu'}  & = \frac{H^2}{8\pi^2} \left( \frac{1}{{\cal Z}-1}g_{\mu\nu'} - n_{\mu} n_{\nu'}  \right)
\\& \\
D^{\sss H}_{\mu+} & = \frac{1}{H} \sqrt{{\cal Z}^2-1} \ n_{\mu}
\\& \\
D^{\sss H}_{++} & =  \frac{1}{8\pi^2},
\end{cases}
\end{equation}
where we have used (de Sitter is a maximally symmetric space) the standard unit tangent vectors $n_{\mu}(x,x')$ and $n_{\nu'}(x,x')$, the parallel propagator along the geodesic $g_{\mu\nu'}(x,x')$ 
and the usual function $\mathcal{Z}$ of the geodesic distance $\mu(x,x')$ relating $x$ and $x'$, $\mathcal{Z}= \cosh\left(H \mu \right)$. See \cite{pconf3} for a more precise statement.
Note that the two-point function $D_{\mu\nu'}^{\sss H}$ has the same physical content with the Allen and Jacobson two-point function \cite{AllenJacobson}.

\section{Conclusion}
An $SO(2,4)$-covariant quantization of the Maxwell field in an arbitrary conformally flat space was presented.
Following Dirac's six-cone formalism, all conformally flat spaces $X_{\sss K}$ are realized as intersections of the null cone and a given surface ${\cal P}_{\sss \tilde{K}}$. 
The quantum field was explicitly constructed using the Gupta-Bleuler canonical quantization scheme and the Wightman two-point functions were given. The price to pay for this simplicity and the maintaining of the $SO(2,4)$ invariance during the whole quantization process was the introduction of two auxiliary fields $A_{c}$ and $A_{+}$. As a consequence, the Maxwell field $A_{\mu}$ does not propagate ``alone'' but together with its two auxiliary fields. The propagation must use all the Wightman functions (\ref{D-X-K-relevent}) and not only the ``purely'' Maxwell one $D_{\mu\nu}$. 
Nonetheless, in a recent work \cite{pconf4}, we have used the functions (\ref{D-ds}) to propagate the Maxwell field generated by two charges of opposite sign placed at the two poles of a de Sitter space. The calculations showed that only $D_{\mu\nu}$ is involved, which trivialize the problem.
One can consider to use the two-point functions (\ref{D-X-K-relevent}) to propagate the electromagnetic field for some charge distribution given in other CFSs, like FLRW spaces for instance.

One concludes that is much worth to maintain the $SO(2,4)$ symmetry during the whole quantization process when dealing with Maxwell field in a conformally flat space. The problem then goes back to Minkowski and the calculations become much easier. In fact, the classical and quantum structures of the free Maxwell field are locally the same in all conformally flat spaces. The remained question is to know if this is true for other free fields and how to deal with $SO(2,4)$-invariant interactions?

\section*{Acknowledgments}
I would like to thank M. Novello, J. Renaud and E. Huguet for illuminating discussions and the CNPq for financial support.

\appendix
\section{The transformations relating $a_{\alpha}$ to $A_{\sss I}$}\label{details}
Considering (\ref{field-a}) and (\ref{eq-1}), expressing the basis $\{dy\}$ in terms of  $\{dx\}$ and then identifying both sides, one obtains the expression of $\tilde{A}_{\sss I}^{\sss M}$in terms of $\tilde a_{\alpha}$. 
We find, after using the homogeneity properties,
\begin{equation}\label{matrix-S}
A_{\sss I}^{\sss M} =  S_{\sss I}^{\sss \beta}(x)\ a_{\alpha}^{\sss M} ,
\end{equation}
which reads
\begin{equation}\label{A(a)-m}
\left \{
 \begin{array}{lcl}
 A_c^{\sss M} &=&   a^{\sss M}_5 - a^{\sss M}_4  
 \\& &\\
 A_\mu^{\sss M} & = &  a^{\sss M}_\mu -  \frac{1}{2}\left( a^{\sss M}_5 - a^{\sss M}_4 \right) x_{\mu} 
 \\ & & \\
 A_+^{\sss M} &=& a^{\sss M}_5 (1- x^2) + a^{\sss M}_4 (1+ x^2) + a^{\sss M}.x.  ~~~~~~~~~~~~~
   \end{array}
\right .
\end{equation} 
This system can be inverted in
\begin{equation}\label{a(A)-m}
\left \{
 \begin{array}{lcl}
{\displaystyle a^{\sss M}_5} &=& {\displaystyle \frac{1}{2} \{ A_{+}^{\sss M}- A^{\sss M}x + A_c^{\sss M} \left(1 - x^2 \right) \} } ~~~~~~~~~~~~~~~~~\\
{\displaystyle a^{\sss M}_4} &=& {\displaystyle \frac{1}{2}
\{ A_{+}^{\sss M} - A^{\sss M}x - A_c^{\sss M} \left(1 +x^2 \right) \} } \\
 {\displaystyle a^{\sss 0}_\mu} &=& {\displaystyle A_{\mu}^{\sss M} +  \frac{1}{2} A_c^{\sss M} x_{\mu} }  .
 \end{array}
\right.
\end{equation}

Following the same steps as above, one obtain the matrix linking the $a_{\alpha}^{\sss K}$ to the $A_{\sss I}^{\sss K}$
\begin{equation}\label{A(a)}
\left \{
 \begin{array}{lcl}
   {\displaystyle  A_c^{\sss K}} 
  &=&   K \biggl\{ a^{\sss K}_5[ \Upsilon_c(1-x^2)+1] 
  + a^{\sss K}_4 [\Upsilon_c(1+x^2)-1] 
  \\&& \qquad \qquad \qquad
  + \Upsilon_c a^{\sss K}.x \biggr\}
\\
   {\displaystyle A_\mu^{\sss K}}
   & = &  K  \biggl\{ a^{\sss K}_5 \left( (1 -x^2) \Upsilon_{\mu} - \frac{1}{2}x_{\mu} \right)  
\\&&  
    +  a^{\sss K}_4 \left( (1+ x^2) \Upsilon_{\mu} + \frac{1}{2} x_{\mu} \right) 
+ a^{\sss K}_\nu \left( \Upsilon_{\mu} x^\nu + \delta_{\mu}^{\nu} \right) \biggr\} \\
 {\displaystyle  A_+^{\sss K}} 
 &=& {\displaystyle K \biggl\{a^{\sss K}_5 (1- x^2) } + a^{\sss K}_4 (1+ x^2) + a^{\sss K}.x \biggr\}.
   \end{array}
\right .
\end{equation} 
This system can be obtained using the Minkowskian system (\ref{A(a)-m}), the relation $a^{\sss K}_{\alpha} = K^{-1} a^{\sss M}_{\alpha}$ (which comes out from the homogeneity properties of the fields) and the extended Weyl transformations (\ref{extendedWeyl-bis}).
This is inverted in
\begin{equation}\label{a(A)}
\left \{
 \begin{array}{lcl}
{\displaystyle a^{\sss K}_5} 
&=& {\displaystyle \frac{1}{2K}
\biggl\{  A_+^{\sss K}  \left(1 + \Upsilon.x - \Upsilon_c(1- x^2) \right) } 
\\&& \qquad \qquad \qquad
- A^{K}x + A_c^{\sss H} \left(1 - x^2 \right) \biggr\} \\
{\displaystyle a^{\sss K}_4} &=& {\displaystyle \frac{1}{2K}
\biggl\{  A_+^{\sss K}  \left(1 +\Upsilon.x + \Upsilon_c(1+ x^2 ) \right) }
\\&& \qquad \qquad \qquad
 -  A^{K}x - A_c^{\sss H} \left(1 + x^2\right) \biggr\}  \\
 {\displaystyle a^{\sss K}_\mu} &=& {\displaystyle \frac{1}{2K}
\left\{ -A_+^{\sss K} \left(2\Upsilon_{\mu}+ \Upsilon_cx_\mu\right) + 2A_{\mu}^{K} + A_c^{\sss K} x_{\mu} \right\}}  .
 \end{array}
\right.
\end{equation}

\section{The undecomposable $SO(2,4)$ representations of the fields $A_{\sss I}$}\label{action}
The $SO(2,4)$ infinitesimal action on the field $A_{\sss I}$ is given by commutators of the group generators  and the field. 
First, we write down the infinitesimal transformations of the Minkowskian fields $A_{\sss I}^{\sss M}$ which can be found in \cite{pconf3} then we transport the resulting representation into $X_{\sss K}$. 
For any element $e\in SO(2,4)$, the related generator is denoted by $X_{e}^{\sss M}$  and whose action on the field  $A_{\sss I}^{\sss M}$ reads
\begin{equation}\label{action-A}
\begin{split}
(X_e^{\sss M}  \ A^{\sss M})_{\sss I} & = 
 [ X_{e}^{\sss M} , A_{\sss I}^{\sss M} ]
\\ & 
= X_{e} \ A_{\sss I}^{\sss M}  + \left( \Sigma_{e}  \right)_{\sss I}^{\sss J} \ A_{\sss J}^{\sss M}
\end{split}
\end{equation}
where the first part represents the scalar action and the second the spinorial action.
Setting
\begin{equation}
\begin{split}
K_{\mu} & = 2  x_{\mu} \ x.\partial - x^2 \ \partial_\mu
\\
 X_{\mu\nu}  & = x_{\mu} \partial_{\nu} - x_{\nu} \partial_{\mu},
\end{split}
\end{equation}
the Minkowskian infinitessimal action reads
\begin{equation*}
\left \{
\begin{aligned}
& \left(X_{\mu\nu}^{\sss M} \ A^{\sss M}\right)_c = 
X_{\mu\nu} A^{\sss M}_c~~~~~~~~~~~~~~~~~~~~~~~~~~~~~~~~~~~~~~~~~~~\\
&
\left(X_{\mu\nu}^{\sss M} \ A^{\sss M}\right)_\lambda =  X_{\mu\nu}A^{\sss M}_\lambda 
 +\eta_{\lambda[\mu} \delta^\tau_{\nu]} \ A^{\sss M}_\tau
 \\ & 
\left( X_{\mu\nu}^{\sss M} \ A^{\sss M}\right)_+ =  X_{\mu\nu} A^{\sss M}_+,
\end{aligned}
\right.
\end{equation*}
for the rotations,
\begin{equation*}
\left \{
\begin{aligned}
&\left(P_\mu^{\sss M} \ A^{\sss M}\right)_c = \partial_\mu  A^{\sss M}_c ~~~~~~~~~~~~~~~~~~~~~~~~~~~~~~~~~~~~~~~~~~~~~~\\
&
\left(P_\mu^{\sss M} \ A^{\sss M}\right)_\nu = \partial_\mu A^{\sss M}_\nu \\ 
& \left(P_\mu^{\sss M} \ A^{\sss M}\right)_+ = \partial_\mu A^{\sss M}_+,
\end{aligned}
\right .
\end{equation*}
for the translations;
\begin{equation*}
\left \{
\begin{aligned}
& \left(K_\mu^{\sss M} \ A^{\sss M}\right)_c = K_\mu  A^{\sss M}_c 
+ 4 (x_{\mu} A_{c}^{\sss M} + A^{\sss M}_\mu)
 \\&
 \left(K_\mu^{\sss M} \ A^{\sss M}\right)_\nu = K_\mu A^{\sss M}_\nu  
 +  2(x_{[\mu} \delta_{\nu]}^\lambda  + x^\lambda \eta_{\mu\nu}) A^{\sss M}_\lambda - 2\eta_{\mu\nu} A^{\sss M}_+  
 \\
&\left(K_\mu^{\sss M} \ A^{\sss M}\right)_+ = K_\mu A^{\sss M}_+,
\end{aligned}
\right.
\end{equation*}
for the special conformal transformations (SCT). Finally, we have
\begin{equation*}
\left \{
\begin{aligned}
&\left(D^{\sss M}\  A^{\sss M}\right)_c = (x\partial + 2) A^{\sss M}_c 
~~~~~~~~~~~~~~~~~~~~~~~~~~~~~~~~~~~~~~~~\\&
\left(D^{\sss M} \ A^{\sss M}\right)_\mu = \left(x\partial + 1\right) A^{\sss M}_\mu 
\\ &
\left(D^{\sss M} \ A^{\sss M}\right)_+ = (x\partial + 0) A^{\sss M}_+,
\end{aligned}
\right .
\end{equation*}
for the dilations.
The undecomposable structure of the fields $A_{\sss I}^{\sss M}$ is made clear. Under the $SO(2,4)$ action, the component $A^{\sss M}_{c}$ overlaps $A^{\sss M}_{\mu}$ which in turn overlaps $A^{\sss M}_{+}$. 
So we have the scheme
\begin{equation}\label{undecomposable}
\left \{
\begin{aligned}
& A^{\sss M}_c \to  A^{\sss M}_+ , A^{\sss M}_\mu ,  A^{\sss M}_c,
~~~~~~~~
\\
&A^{\sss M}_\mu \to  A^{\sss M}_+, A^{\sss M}_\mu , 
\\ &
 A^{\sss M}_+ \to  A^{\sss M}_+.
\end{aligned}
\right .
\end{equation}

The second step is to trasport the group action from Minkowski to the $X_{K}$ space using the extended Weyl transformation (\ref{extendedWeyl-bis})
\begin{equation}\label{transport-weyl}
\begin{split}
(X_e^{\sss K}  \ A^{\sss K})_{\sss I} 
& = W_{\sss I}^{\sss J} \ (X_e^{\sss M}  \ A^{\sss M})_{\sss J} 
\\
& = (X_e^{\sss M}  \ A^{\sss M})_{\sss I} + \Upsilon_{\sss I} (X_e^{\sss M}  \ A^{\sss M})_{+}
\\ & 
=   X_{e} \ A_{I}^{\sss M}  + \left(  \Sigma_{e}  \right)_{\sss I}^{\sss J}  \, A^{\sss M}_{\sss J} + \Upsilon_{\sss I} X_{e} \ A_{+}^{\sss M}
\\ &
=   X_{e} \ (A_{I}^{\sss K} - \Upsilon_{\sss I} A_{+}^{\sss K} )  + \left(  \Sigma_{e} \right)_{\sss I}^{\sss J}  \, (A_{J}^{\sss K} - \Upsilon_{\sss J} A_{+}^{\sss K} )  
\\ & \qquad \qquad + \Upsilon_{\sss I} X_{e} \ A_{+}^{\sss K}
 \\&
= (X_{e}^{M} \ A^{\sss K})_{I} - \left[ (X_{e}\Upsilon_{I}) + \left(\Sigma_{e} \right)_{I}^{J} \Upsilon_{J}  \right] A_{+}^{\sss K}
\end{split}
\end{equation}
where we have used $\left(\Sigma_{e} \right)_{+}^{J} = 0$ for all $e \in SO(2,4)$. 
Also only the second part of the last line has to be computed.

The infinitesimal $SO(2,4)$ action on $A_{\sss I}^{\sss K}$ reads
\begin{equation*}
\left \{
\begin{aligned}
& \left(X_{\mu\nu}^{\sss K} \ A^{\sss K}\right)_c = 
X_{\mu\nu} A^{\sss K}_c - (X_{\mu\nu}\Upsilon_{c})A_{+}^{\sss K}~~~~~~~~~~~~~~~~~~~~~~~~~~~~~~~~~~~~~~~~~\\
&
\left(X_{\mu\nu}^{\sss K} \ A^{\sss K}\right)_\lambda =  X_{\mu\nu}A^{\sss K}_\lambda 
 + \eta_{\lambda[\mu} \delta^\tau_{\nu]}  \ A^{\sss K}_\tau  
\\& \qquad \qquad \qquad
 - \left[ (X_{\mu\nu}\Upsilon_{\lambda}) + \eta_{\lambda[\mu} \delta^\tau_{\nu]} \Upsilon_{\tau}  \right] A_{+}^{\sss K}
 \\ & 
\left( X_{\mu\nu}^{\sss K} \ A^{\sss K}\right)_+ =  X_{\mu\nu} A^{\sss K}_+,
\end{aligned}
\right.
\end{equation*}
for the rotations,
\begin{equation*}
\left \{
\begin{aligned}
&\left(P_\mu^{\sss K} \ A^{\sss K}\right)_c = \partial_\mu  A^{\sss K}_c - (\partial_{\mu} \Upsilon_{c}) A_{+}^{\sss K} ~~~~~~~~~~~~~~~~~~~~~~~~~~~~~~~~~~~~~~~~~~~~~~\\
&
\left(P_\mu^{\sss K} \ A^{\sss K}\right)_\nu = \partial_\mu A^{\sss K}_\nu - (\partial_{\mu} \Upsilon_{\nu}) A_{+}^{\sss K} 
\\ 
& \left(P_\mu^{\sss K} \ A^{\sss K}\right)_+ = \partial_\mu A^{\sss K}_+ .
\end{aligned}
\right .
\end{equation*}
for the translations and
\begin{equation*}
\left \{
\begin{aligned}
& \left(K_\mu^{\sss K} \ A^{\sss K}\right)_c = K_\mu  A^{\sss K}_c 
+ 4 (x_{\mu} A_{c}^{\sss K} + A^{\sss K}_\mu) 
\\& \qquad \qquad \qquad
-  \left[ (K_{\mu}\Upsilon_{c}) + 4( x_{\mu}\Upsilon_{c} + \Upsilon_{\mu} ) \right] A_{+}^{\sss K}
 \\&
 \left(K_\mu^{\sss K} \ A^{\sss K}\right)_\nu = K_\mu A^{\sss K}_\nu  
 +  2(x_{[\mu} \delta_{\nu]}^\lambda  + x^\lambda \eta_{\mu\nu}) A^{\sss K}_\lambda - 2\eta_{\mu\nu} A^{\sss K}_+  
 \\& \qquad \qquad \qquad
 -  \left[ (K_{\mu}\Upsilon_{\nu}) +  2(x_{[\mu} \delta_{\nu]}^\lambda  + x^\lambda \eta_{\mu\nu}) \Upsilon_{\lambda} \right] A_{+}^{\sss K}
 \\
&\left(K_\mu^{\sss K} \ A^{\sss K}\right)_+ = K_\mu A^{\sss K}_+,
\end{aligned}
\right.
\end{equation*}
for the special conformal transformations. Finally, we have
\begin{equation*}
\left \{
\begin{aligned}
&\left(D^{\sss K}\  A^{\sss K}\right)_c = (x\partial + 2) A^{\sss K}_c 
-  \left[ (x.\partial \Upsilon_{c}) +  2  \Upsilon_{c})  \right] A_{+}^{\sss K}
~~~~~~~~~~~~~~~~~~~~~~~~\\&
\left(D^{\sss K} \ A^{\sss K}\right)_\mu = \left(x\partial + 1\right) A^{\sss K}_\mu 
-  \left[ (x.\partial \Upsilon_{\mu}) +   \Upsilon_{\mu})  \right] A_{+}^{\sss K}
\\ &
\left(D^{\sss K} \ A^{\sss K}\right)_+ = (x\partial + 0) A^{\sss K}_+,
\end{aligned}
\right .
\end{equation*}
for the dilations. The undecomposable structure of the fields $A_{\sss I}^{\sss K}$ appears in a similar way than  for the Minkowskian fields (\ref{undecomposable}).


Note that the constraint $A^{\sss M}_{+}=A^{\sss K}_{+}=0$ (\ref{A+=0}) is $SO(2,4)$-invariant. This is important since this constraint defines the subset of physical states.

\section{The scalar product}\label{PS}
The $SO(2,4)$-invariant scalar product for the Minkowskian field $A_{\sss I}^{\sss M}$ reads
\begin{equation}\label{scalarPS-A}
\begin{split}
\langle A^{\sss M}, B^{\sss M}\rangle
&=  - i\, \int_{\Sigma}\!\!\!\sigma^\mu_{\sss M}\, J_{\mu}(A^{\sss M}, B^{\sss M})
\\&  = - i\, \int_{\Sigma}\!\!\!\sigma^\mu_{\sss M}\,
\Bigl\{ A^{*\nu}_{\sss M}\stackrel{\leftrightarrow}{\partial_\mu}B_\nu^{\sss M}
+ ( A_\mu^{{\sss M}*}B_c^{\sss M} - A_c^{{\sss M}*}B_\mu^{\sss M} ) 
\\& \qquad \qquad \qquad
 + \frac{1}{2} ( A_+^{{\sss M}*}\stackrel{\leftrightarrow}{\partial_\mu}B_c^{\sss M}
+  A_c^{{\sss M}*} \stackrel{\leftrightarrow}{\partial_\mu}B_+^{\sss M})
\Bigr \},
\end{split}
\end{equation}
where $\Sigma$ is some Cauchy surface in $X_{\sss M}$ and $\sigma^\mu_{\sss M}$ is a surface element. An important point is that this Cauchy surface is common to all the spaces $X_{\sss K}$ since they are all conformally equivalent \cite{Fulling}.


Using (\ref{lien-weyl}) and (\ref{extendedWeyl-bis}), the scalar product for the field $A_{\sss I}^{\sss K}$ is obtained from (\ref{scalarPS-A}) and reads
\begin{equation}\label{scalarPS-A-K}
\begin{split}
\langle A^{\sss K}, B^{\sss K}\rangle
& = \langle A^{\sss M}, B^{\sss M}\rangle 
\\&=  - i\, \int_{\Sigma}\!\!\!\sigma^\mu_{\sss M} \ J_{\mu}(W^{-1} A^{\sss K}, W^{-1} B^{\sss K})
\\&  = - i\, \int_{\Sigma}\!\!\!\sigma^\mu_{\sss K}\,
\Bigl\{ (A^{*\nu}_{\sss K} -\Upsilon^{\nu} A_{+}^* )\stackrel{\leftrightarrow}{\partial_\mu} (B_\nu^{\sss M} - \Upsilon_{\nu} B_{+}) 
\\& \qquad \qquad 
 + \frac{1}{K^{2}}  (A_\mu^{{\sss K}*}- \Upsilon_{\mu} A_{+}^*) (B_c^{\sss K} - \Upsilon_{c}A_{+}) 
\\&  \qquad \qquad  - \frac{1}{K^{2}}  ( A_c^{{\sss M}*} - \Upsilon_c A_{+}^* ) (B_\mu^{\sss M} - \Upsilon_\mu A_{+}) 
\\& \qquad \qquad 
  + \frac{1}{2 K^{2}} ( A_+^{{\sss M}*}\stackrel{\leftrightarrow}{\partial_\mu} (B_c^{\sss M} - \Upsilon_{c}A_{+}) 
\\&  \qquad \qquad  + \frac{1}{2K^{2}}  ( A_c^{{\sss M}*} - \Upsilon_c A_{+}^* )  \stackrel{\leftrightarrow}{\partial_\mu}B_+^{\sss M}),
\Bigr \}
\end{split}
\end{equation}
where the $X_{\sss K}$ surface element is related to its Minkoskian counterpart by $\sigma^\mu_{\sss K}=K^2\, \sigma^\mu_{\sss M}$. 
%




\end{document}